\documentclass{PoS}

\title{Chiral symmetry restoration and the thermal $f_0(500)$ state.}

\ShortTitle{Chiral symmetry restoration and the thermal $f_0(500)$ state.}

\author{\speaker{Andrea Vioque-Rodr\'iguez}\\
        Departamento de F\'isica Te\'orica and IPARCOS. Univ. Complutense. 28040 Madrid, Spain\\
        E-mail: \email{avioque@ucm.es}}

\author{Angel G\'omez Nicola\\
        Departamento de F\'isica Te\'orica and IPARCOS. Univ. Complutense. 28040 Madrid, Spain\\
        E-mail: \email{gomez@ucm.es}}

\abstract{

We analize the role played by the thermal $f_0(500)$ state or $\sigma$ in chiral symmetry restoration. The temperature corrections to the spectral properties of that state are included in order to provide a better description of the scalar susceptibility $\chi_S$ around the transition region. We use the Linear Sigma Model to establish the relation between $\chi_S$ and the $\sigma$ propagator, which is used as a benchmark to test the approach where $\chi_S$ is saturated by the $f_0(500)$ inverse self-energy. Within such saturation approach, a peak for $\chi_S$ around the chiral transition is obtained when considering the $f_0(500)$ generated as a $\pi\pi$ scattering pole within Unitarized Chiral Perturbation Theory at finite temperature. That approach yields results complying with lattice data when the uncertainties of the low-energy constants are taken into account. Those uncertainties and the unitarization method are used to check the robustness of this approximation. Finally, we will discuss some recent results within the chiral lagrangian framework related to the topological susceptibility and its connection with chiral and $U_A(1)$ restoration.
}

\FullConference{Light Cone 2019 - QCD on the light cone: from hadrons to heavy ions - LC2019\\
		16-20 September 2019\\
		Ecole Polytechnique, Palaiseau, France}

\begin{document}

\section{Introduction}

On the one hand these proceedings review our recent analyses of the role of the f0(500) state in chiral symmetry restoration \cite{Ferreres-Sole:2018djq}. For vanishing barion density and $ 2 + 1$ flavours with $m_u=m_d=m_l\ll m_s$ quark masses the chiral transition is a crossover at a transition temperature of about $T_c\sim 156$ MeV \cite{Bazavov:2011nk,Aoki:2009sc}. The light quark condensate $\langle \bar{q}q\rangle_l=\langle\bar{u}u+\bar{d}d\rangle$ and the scalar susceptibility $\chi_S$ are the most used parameters to study the restoration of chiral symmetry.
In \cite{Ferreres-Sole:2018djq} we show the relation between $\chi_S$ and the sigma self-energy in the Linear Sigma Model (LSM). In section 2, we will study the dependence on the coupling constant of the model and the momentum of the sigma self-energy at finite temperature. Next, we will consider the $\sigma$ resonance which can be generated within Unitarized Chiral Perturbation Theory (UChPT) and we will present a saturated approach which will be tested using different analyses.

On the other hand we have calculated the topological susceptibility and the forth order cumulant up to NNLO in U(3) ChPT including the leading isospin-breaking corrections \cite{Nicola:2019ohb}. From that the effect of the additional U(3) corrections in terms of the Low Energy Constants (LECs) will be estimated. Finally, we will extend our analysis to finite temperature to analyze the Ward Identity which relates $\langle\bar{q}q\rangle_l$ with the topological susceptibility ($\chi_{top}$) and the pseudoscalar susceptibility in the light $\eta_l=i\bar u\gamma_5u+i\bar d\gamma_5d$ channel ($\chi_P^{ll}$). An important issue studied is if the term proportional to $\chi_P^{ll}$ is suppressed with respect to $\langle\bar{q}q\rangle_l$.

\section{Linear Sigma Model description of the scalar susceptibility and the thermal $f_0(500)$ saturation approach}

First we consider the light meson sector of the Linear Sigma Model lagrangian, since it includes explicitly the scalar $\sigma$ field:
\begin{equation}
{\cal L}_{LSM}= \frac{1}{2}\left(\partial_\mu\sigma\partial^\mu\sigma+\partial_\mu\pi^a\partial^\mu\pi^a\right)-\frac{\lambda}{4}\left[\sigma^2+\pi_a\pi^a-v_0^2\right]^2+h\sigma.
\label{lsm1}
\end{equation}

We proceed defining a shifted sigma field in such a way that the new field is $\tilde{\sigma}=\sigma-v$, where $v$ is the $\sigma$ expectation value to leading order. Doing that, we have to take into account that one-particle reducible diagrams enter in the calculation of the $\tilde{\sigma}$ propagator. It is because $\langle\tilde{\sigma}\rangle=v(T)-v\neq 0$ at finite temperature. We obtain that $\chi_S$ calculated withing the LSM lagrangian is:
\begin{equation}
\chi_S (T)=-\frac{\partial}{\partial m_l}\langle\bar{q}q\rangle_l=\left(\frac{d^2 h}{d m_l^2}\right)v(T)+\left(\frac{d h}{d m_l}\right)^2 \Delta_\sigma (k=0;T),
\label{sus}
\end{equation}
where $\Delta_\sigma (k;T)=1/[k^2+M_{0\sigma}^2+\Sigma(k_0,\vec{k};T)]$ is the Euclidean propagator of the $\tilde\sigma$ field and $\Sigma(k_0,\vec{k};T)$ is the self-energy.

Near the transition, the term proportional to $v(T)$ in (\ref{sus}) is expected to be negligible because of the quark condensate behaviour. That result can be reached using the Ward Identity $\chi_{\pi}=-\langle\bar{q}q\rangle_l/m_l$ \cite{Buchoff:2013nra,Nicola:2013vma} which implies, together with the equivalence $\chi_{\pi}\simeq\chi_{S}$ around the transition region, that this term is $\mathcal{O}(M_{0\pi}^2/M_{0\sigma}^2)$ suppressed. Thus, the temperature dependence of the susceptibility is approximately described by
\begin{equation}
\frac{\chi_S (T)}{\chi_S (0)}\simeq \frac{M_{0\sigma}^2+\Sigma\left(k=0;T=0\right)}{M_{0\sigma}^2+\Sigma\left(k=0;T\right)}.
\label{susgreen}
\end{equation}

We have extended the calculation of the $\sigma$ self-energy, given in \cite{Pelaez:2015qba,Masjuan:2008cp} and \cite{Ayala:2000px} at zero temperature and a finite temperature respectively, out of the chiral limit. Calculating perturbatively the pole of the propagator and comparing that with the pole of a Breit-Wigner resonance $s_p=(M_p-i\Gamma_p/2)^2$ we obtain that to achieve reasonable phenomenological results the $\lambda$ parameter has to be large. Moreover, it is not possible to get good agreement both for $M_p$ and $\Gamma_p$ but we have selected a reference range $\lambda\sim 10-20$ for which the experimental determination of the mass (width) is recovered with the lower (higher) value of $\lambda$.
\begin{figure}
	\centering
		\includegraphics[width=7cm]{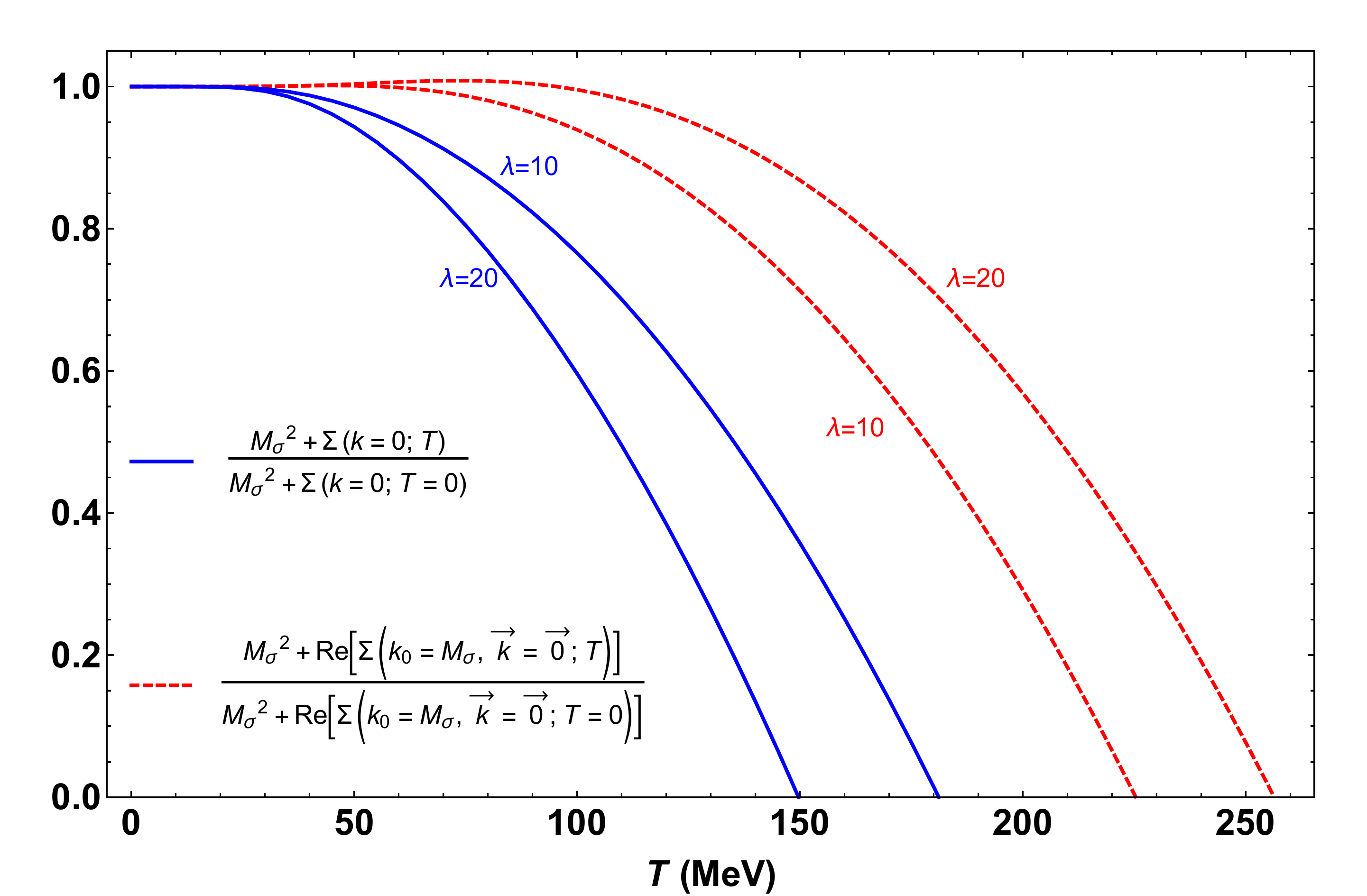}
		\includegraphics[width=7cm]{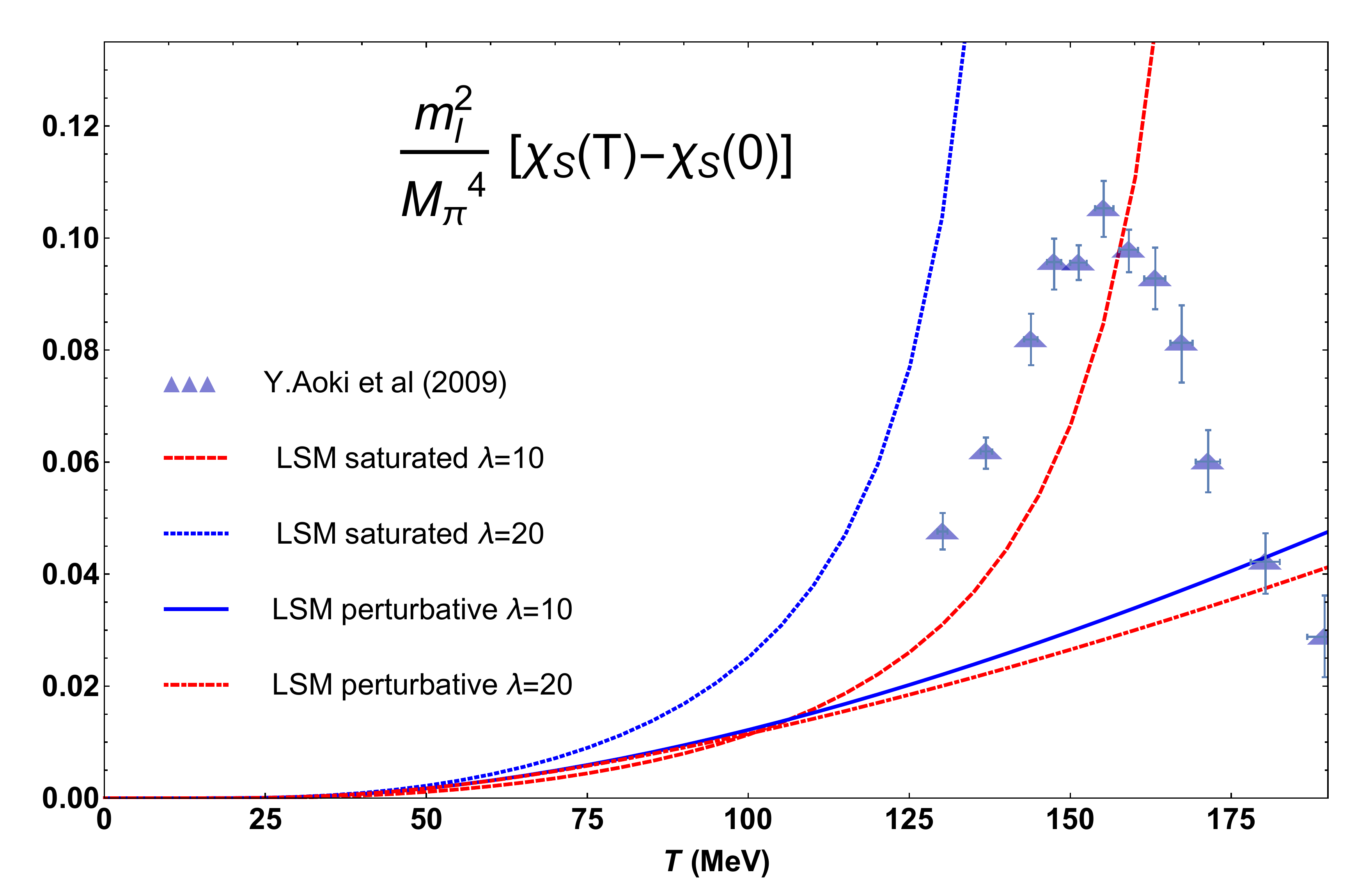}
	\caption{Left: Comparison of the sigma self-energy at $k=0$ and at $k_0=M_\sigma$, $\vec{k}=\vec{0}$, for $\lambda=10-20$. Right: Saturated and perturbatively susceptibility in the LSM compared to lattice data from \cite{Aoki:2009sc}. }
	\label{fig:suscomplsm}
\end{figure}

In Figure \ref{fig:suscomplsm}, we show the temperature dependence of the real part of the self-energy evaluated at $s=0$ and $s=s_p$ \cite{Ferreres-Sole:2018djq}. Although both functions have the same qualitative behaviour the first goes to zero at a lower temperature. In respect of the saturated susceptibility, it diverges around $T_c$ and presents a chiral symmetry restoration tendency but is not able to reproduce the crossover peak. Despite the limitations of the LSM, (\ref{susgreen}) describes lattice data reasonably below $T_c$.

It is well established that the $f_0$(500) is generated within UChPT as a second
Riemann sheet pole of the $\pi\pi$ scattering amplitude. Thus, the pole position parameters of the $f_0$(500) agree with their experimental values. The square of the scalar thermal pole mass, $M_S^2(T)$ is deﬁned as the real part of the self-energy.
The theoretical uncertainties involved in $M_S(T)$ are the unitarization method and the numerical uncertainties of the Low Energy Constants (LECs).
\begin{figure}
\centering
\includegraphics[width=7cm]{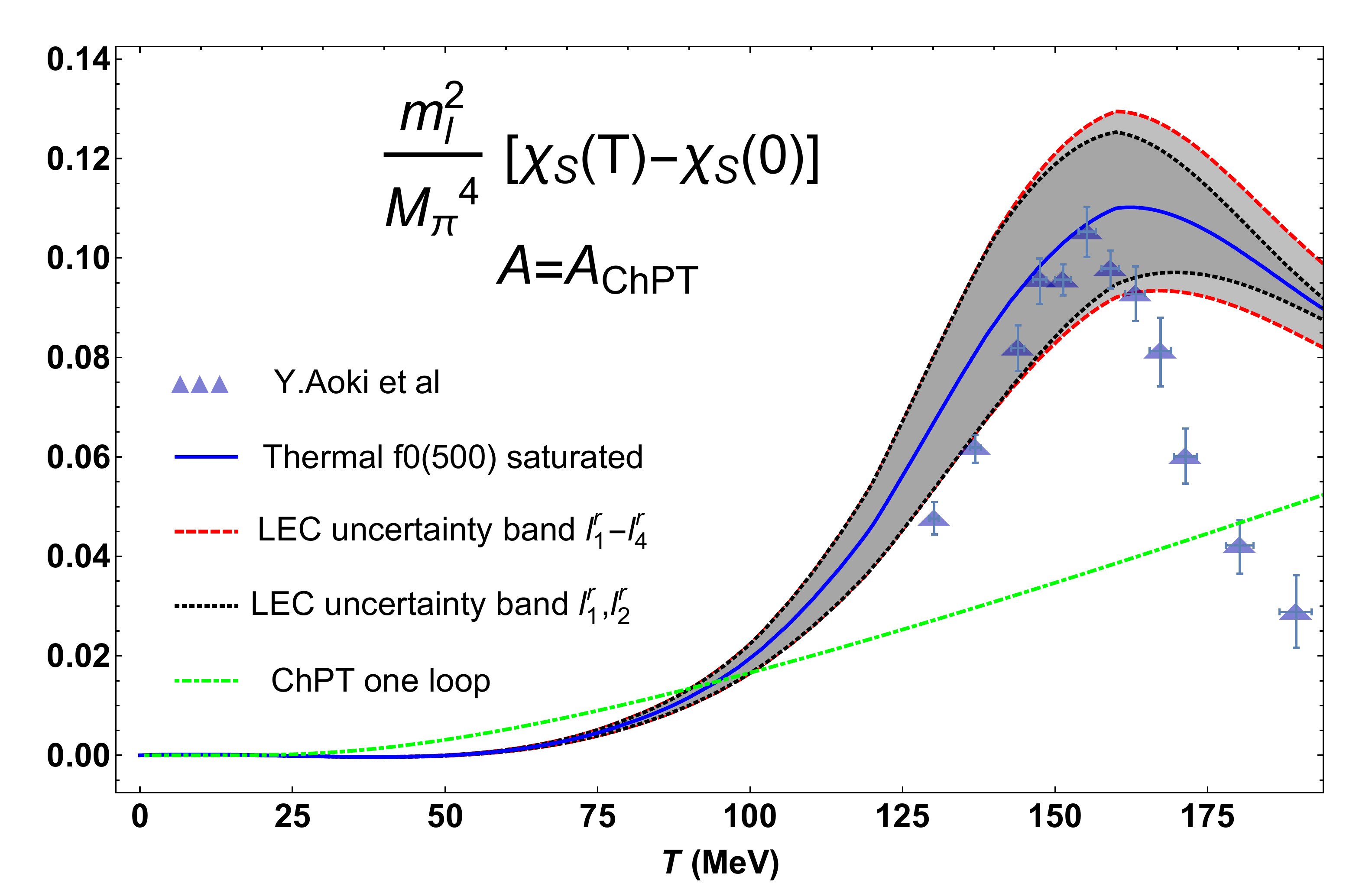}
\caption{Saturated susceptibility  normalized with $A_{ChPT}\simeq 0.15$ including the uncertainties coming from the LEC given in \cite{Hanhart:2008mx}.}
\label{fig:unitsuslec}
\end{figure}

If the momentum dependence of the real part of the self-energy is soft in such a way that the difference between Re$\Sigma_{f_0}$ from $s=0$ to $s=s_p$ lies into the uncertainty range of the approach, we can define the corresponding saturated susceptibility within UChPT which we denote unitarized susceptibility:
\begin{equation}
\chi_S^U(T)=A\frac{M_\pi^4}{4m_l^2}\frac{M_S^2(0)}{M_S^2(T)}.
\label{susunit}
\end{equation}

As we can see in figure \ref{fig:unitsuslec}, the Inverse Amplitude Method (IAM) prediction for the susceptibility reproduces the crossover peak \cite{Nicola:2019umf} and most of the lattice data fall into the uncertainty band below $T_c$. A comparison with the Hadron Resonance
Gas (HRG) is studied in this context in \cite{Ferreres-Sole:2018djq}.

\section{The topological susceptibility in U(3) Chiral Perturbation Theory}

The topological susceptibility and the fourth order cumulant to LO, in terms of quark masses and considering the isospin-breaking corrections, are given by
\begin{equation}
\chi_{top}^{U(3),LO}= \Sigma {\hat m},\hspace{2cm}c_4^{U(3),LO}=-\Sigma \frac{{\hat m}^4}{{\bar m}^{[3]}},
\label{chitoploib}
\end{equation}
with
\begin{equation}
{\hat m}=\frac{M_0^2 {\bar m}}{M_0^2+6B_0{\bar m}}, \hspace{1cm}\bar{m}=\frac{1}{m_u}+\frac{1}{m_d}+\frac{1}{m_s},\hspace{1cm}{\bar m}^{[3]}= \left[\frac{1}{m_u^3}+\frac{1}{m_d^3}+\frac{1}{m_s^3}\right]^{-1}.
\end{equation}

We have calculated up to NNLO corrections in U(3) ChPT \cite{Nicola:2019ohb}, where the $\eta’$ is considered as a ninth Godstone boson within the large $N_c$ framework, including also the relevant isospin-breaking and finite-temperature corrections. Numerical values for the LEC involved and their uncertaintites are taken from \cite{Guo:2015xva}.
The numerical results, which can be seen in \cite{Nicola:2019ohb} in the isospin limit, have been calculated keeping the numerical values of the U(3) LECs and their uncertainties. We get that the $\eta'$ meson and mixing angle corrections are comparable to the kaon and $\eta$ ones introduced in the SU(3) approach. Furthermore, this calculation is compatible with the lattice results within the range of uncertainty. Numerical corrections due to isospin breaking remain below the $5\%$ level.
\begin{figure}
\centering
	\includegraphics[width=7cm]{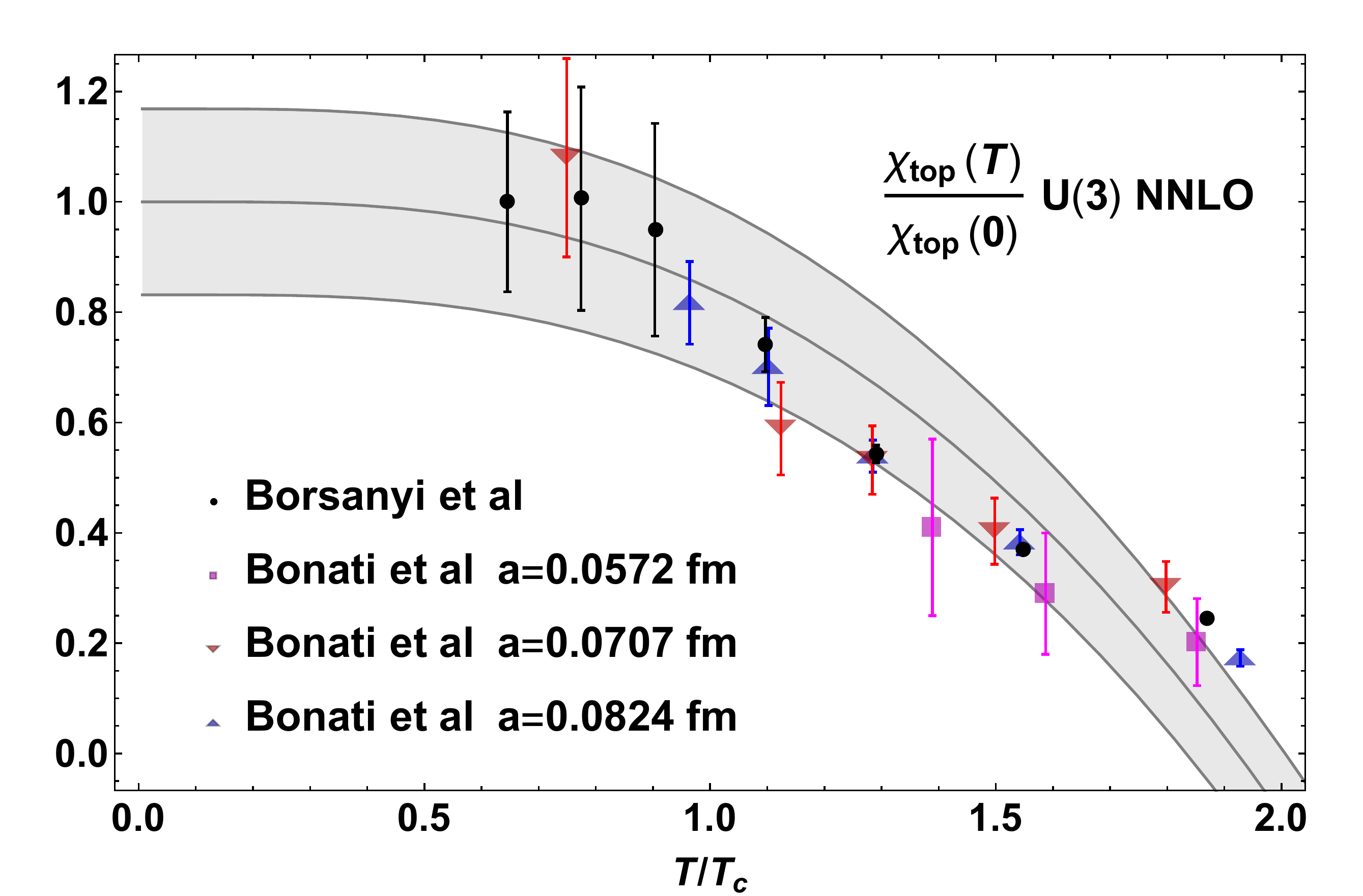}
	\includegraphics[width=7cm]{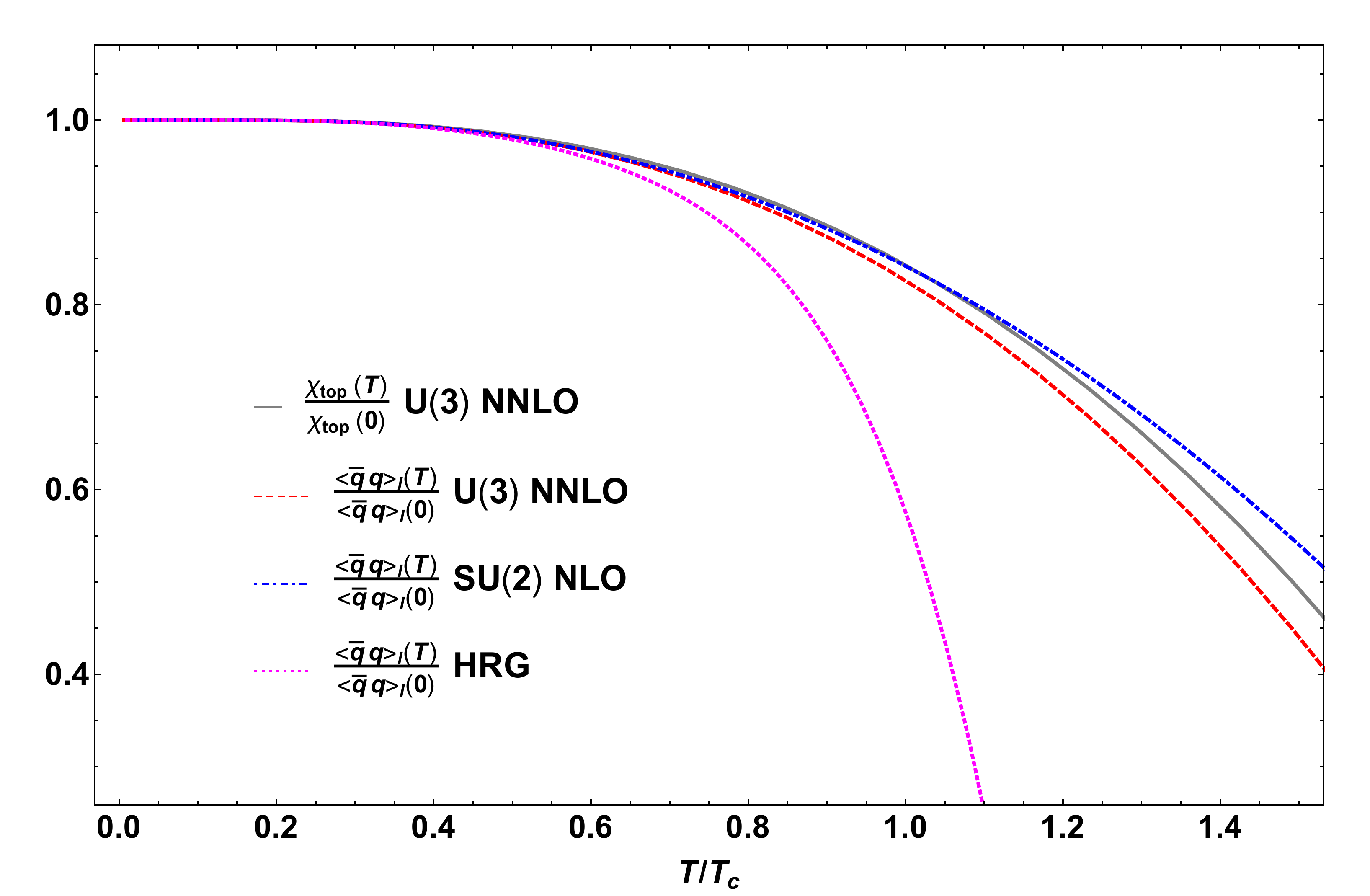}
\caption{Left: Comparison of $\chi_{top}(T)$ calculated to NNLO in U(3) with lattice data from \cite{Bonati:2015vqz} and \cite{Borsanyi:2016ksw}. Right: Topological susceptibility scaling compared to different approaches of the $\langle\bar q q\rangle_l$ scaling.}
\label{fig:temp1}
\end{figure}

In figure 3 we show the temperature dependence of $\chi_{top}$, coming from the meson loops. We can see the consistency between the U(3) ChPT analysis and lattice data within the theoretical and lattice uncertainties. Beside this another important result is that, around the transition point, the $\chi_P^{ll}$ contribution in the Ward identity $\chi_{top}=-m_l(\langle\bar q q\rangle_l+m_l\chi_P^{ll})/4$ \cite{Nicola:2018vug}, may be significant, implying a departure of the scaling of $\chi_{top}$ with the light quark condensate. Thus, in figure 3 we plot the comparison with the quark condensate calculated from the HRG \cite{Jankowski:2012ms} which clearly deviates from $\chi_{top}$ (dominated by the light degrees of freedom) around $T_c$. The latter is consistent with a gap between chiral and $U(1)_A$ restoration for physical quark masses.

\section{Conclusions}

We have related the scalar susceptibility in the LSM to the propagator of the lightest scalar state at zero momentum. After this, from the UChPT approach and using the IAM we have reproduced the crossover peak and have found that most of the lattice data fall into the uncertainty band. Lastly, as regards $\chi_S$, the analyses (LEC, unitarization method, LSM) confirms that the thermal $f_0(500)$ saturation approach is robust from the theoretical point of view. And finally, we have estimated the $\eta'$ corrections to $\chi_{top}$ and $c_4$ at zero temperature and at finite temperature up to NNLO in ChPT. \\

\textit{Acknowledgements} Work partially supported by research contract FPA2016-75654-C2-2-P (spanish ``Ministerio de Econom\'ia y Competitividad''). This work has also received funding from the European Union Horizon 2020 research and innovation programme under grant agreement No 824093. A. V-R acknowledges support from a fellowship of the UCM predoctoral program.


\begin{thebibliography}{99}


\bibitem{Ferreres-Sole:2018djq}
  S.~Ferreres-Sol\'e, A.~G\'omez Nicola and A.~Vioque-Rodr\'iguez,
  Phys.\ Rev.\ D {\bf 99} (2019) no.3,  036018.
  

\bibitem{Bazavov:2011nk}
A.~Bazavov {\it et al.},
Phys.\ Rev.\ D {\bf 85} (2012) 054503.


\bibitem{Aoki:2009sc}
Y.~Aoki, S.~Borsanyi, S.~Durr, Z.~Fodor, S.~D.~Katz, S.~Krieg and K.~K.~Szabo,
JHEP {\bf 0906} (2009) 088.

\bibitem{Nicola:2019ohb}
  A.~G\'omez Nicola, J.~Ruiz De Elvira and A.~Vioque-Rodr\'iguez,
  JHEP {\bf 1911} (2019) 086.
  
\bibitem{Buchoff:2013nra}
M.~I.~Buchoff {\it et al.},
Phys.\ Rev.\ D {\bf 89} (2014) no.5,  054514.

\bibitem{Nicola:2013vma}
A.~G\'omez Nicola, J.~Ruiz de Elvira and R.~Torres Andres,
Phys.\ Rev.\ D {\bf 88} (2013) 076007.


\bibitem{Pelaez:2015qba}
J.~R.~Pelaez,
Phys.\ Rept.\  {\bf 658} (2016) 1.

 
 \bibitem{Masjuan:2008cp}
P.~Masjuan, J.~J.~Sanz-Cillero and J.~Virto,
Phys.\ Lett.\ B {\bf 668} (2008) 14.


\bibitem{Ayala:2000px}
A.~Ayala and S.~Sahu,
Phys.\ Rev.\ D {\bf 62} (2000) 056007.


\bibitem{Hanhart:2008mx}
C.~Hanhart, J.~R.~Pelaez and G.~Rios,
Phys.\ Rev.\ Lett.\  {\bf 100} (2008) 152001.


\bibitem{Nicola:2019umf}
  A.~G.~Nicola, J.~Ruiz de Elvira and A.~Vioque-Rodr\'iguez,
  arXiv:1910.09649 [hep-ph].


\bibitem{Guo:2015xva}
  X.~K.~Guo, Z.~H.~Guo, J.~A.~Oller and J.~J.~Sanz-Cillero,
  JHEP {\bf 1506} (2015) 175.
  
 
  \bibitem{Bonati:2015vqz}
  C.~Bonati, M.~D'Elia, M.~Mariti, G.~Martinelli, M.~Mesiti, F.~Negro, F.~Sanfilippo and G.~Villadoro,
  JHEP {\bf 1603} (2016) 155.
  
 \bibitem{Borsanyi:2016ksw}
  S.~Borsanyi {\it et al.},
  Nature {\bf 539} (2016) no.7627,  69.
  
\bibitem{Nicola:2018vug}
  A.~G\'omez Nicola and J.~Ruiz De Elvira,
  Phys.\ Rev.\ D {\bf 98} (2018) no.1,  014020.
  
\bibitem{Jankowski:2012ms}
  J.~Jankowski, D.~Blaschke and M.~Spalinski,
  Phys.\ Rev.\ D {\bf 87} (2013) no.10,  105018.
 


\end{thebibliography}
\end{document}